%% file: ang.tex
\documentclass{llncs}

\usepackage{amssymb}
\usepackage{amsmath}
\usepackage{epsfig}
\usepackage{float}

\newcommand{\be}{\begin{enumerate}}
\newcommand{\ee}{\end{enumerate}}

\author{Thomas Ang \and Jeffrey Shallit}
\institute{David R.\ Cheriton School of Computer Science,\\
University of Waterloo,
Waterloo, ON, Canada N2L 3G1\\
\email{
\{tang,shallit\}@uwaterloo.ca} }

\title{Length of the Shortest Word in the Intersection of Regular Languages}

\begin{document}

\maketitle

\begin{abstract}
In this note,
we give a construction that provides a tight lower bound of $mn-1$ for the length of the shortest word in the intersection of two regular languages with state complexities $m$ and $n$. 
\end{abstract}

\section{Introduction}

Maslov observed that the state complexity of the intersection of two
regular languages that have state complexities $m$ and $n$ has an upper
bound of $mn$ \cite{Maslov}. One can easily verify this result using the
usual cross-product construction \cite[p.\ 59]{Hopcroft&Ullman:1979}.
This means that
the shortest word in such an intersection cannot be longer than $mn-1$.
It is natural to wonder if this bound is the best possible, over a
fixed alphabet size, for every
choice of $m$ and $n$. Here we show that
there is a matching lower bound.

First we define some notation. A deterministic finite automaton (DFA)
is a quintuple $(Q, \Sigma, \delta, q_0, A)$ where $Q$ is
the finite set of states, $\Sigma$ is the finite input alphabet,
$\delta:Q\times\Sigma\rightarrow Q$ is the transition function, $q_0
\in Q$ is the initial state, and $A \subseteq Q$ is the set of
accepting states. For a DFA $M$, $L(M)$ denotes the language accepted by
$M$. For any $x \in \Sigma^*$, $|x|$ denotes the length of $x$, and
$|x|_a$ for some $a \in \Sigma$ denotes the number of occurrences of
$a$ in $x$. We also define two maps from nonempty languages to
${\mathbb N}$ as follows. For a nonempty language $L$, let ${\rm
lss}(L)$ denote the length of the shortest word in $L$. If $L$ is
regular, then we let ${\rm sc}(L)$ denote the state complexity of $L$
(the minimal number of states in any DFA accepting $L$).

We previously stated that the upper bound on the state complexity of
the intersection of two regular languages implies an upper bound the
length of the shortest word in the intersection. More precisely, we
have ${\rm lss}(L) < {\rm sc}(L)$, which follows
directly from the pumping lemma for regular languages
\cite[p.\ 55]{Hopcroft&Ullman:1979}.
So all that is left is to show that the upper
bound of $mn-1$ can actually be attained for all $m$ and $n$. There is
an obvious construction over a unary alphabet that works when
$\gcd(m,n) = 1$: namely, set
	\begin{itemize}
	\item $L_1 = \{ x : |x| \equiv m-1 \pmod m\}$, and
	\item $L_2 = \{ x : |x| \equiv n-1 \pmod n\}$.
	\end{itemize}
However, this construction
fails when $\gcd(m,n) \neq 1$, so we provide a more general
construction over a binary alphabet that works for all $m$ and $n$.

\section{Our result}

\begin{proposition}
\label{prop:basic}
For all integers $m, n \geq 1$ there exist DFAs $M_1, M_2$ with $m$ and $n$ states, respectively, such that $L(M_1) \cap L(M_2) \not= \emptyset$, and ${\rm lss}(L(M_1) \cap L(M_2) ) = mn-1$. 
\end{proposition}
\proof
The proof is constructive. Without loss of generality, assume $m \leq
n$, and set $\Sigma = \{0,1\}$. Let $M_1$ be the DFA given
by $(Q_1, \Sigma,
\delta_1, p_0, A_1)$, where $Q_1 = \{p_0, p_1, p_2,\ldots, p_{m-1}\}$,
$A_1 = p_0$, and for each $a$, $0 \leq a \leq m-1$, and 
$c \in \lbrace 0, 1 \rbrace $ we set
\begin{equation}
\label{eq:delta1}
\delta_1(p_a, c) = p_{(a+c) \bmod m}.
\end{equation}
Then $$L(M_1) = \{ x \in \Sigma^* : |x|_1 \equiv 0\!\! \pmod {m} \}.$$ 

\begin{figure}[H]
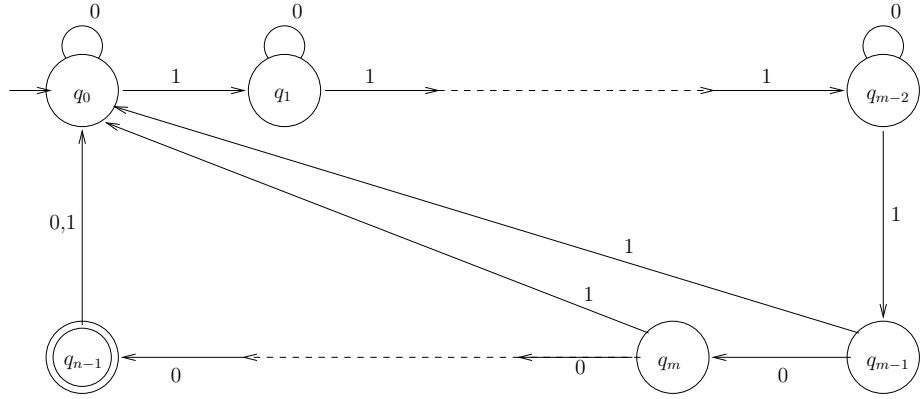

\begin{center}
\resizebox{\columnwidth}{!}{\input intersectm3.pstex_t}
\end{center}

\caption{
	The DFA $M_2$. 
}
\label{fig:m2}
\end{figure}

Let $M_2$ be the DFA $(Q_2, \Sigma, \delta_2, q_0, A_2)$, illustrated in Figure \ref{fig:m2}, where $Q_2 = \{q_0, q_1, q_2,\ldots, q_{n-1}\}$, $A_2 = q_{n-1}$, and for each $a$, $0 \leq a \leq n-1$,
\begin{equation}
\label{eq:delta2}
\delta_2(q_a, c) =
\begin{cases}
q_{a+c}, & \text{if } 0 \leq a < m-1; \\
q_{(a+1) \bmod n}, & \text{if }c = 0 \text{ and } m-1 \leq a \leq n-1; \\
q_0, & \text{if }c = 1 \text{ and } m-1 \leq a \leq n-1.
\end{cases} \nonumber
\end{equation}

Focussing solely on the $1$'s that appear in some accepting path in
$M_2$, we see that we can return to $q_0$ 
\begin{itemize}
\item[(a)] via a simple path with $m$ $1$'s, or
\item[(b)] (if we go through $q_{n-1}$), via a simple path with 
$m-1$ $1$'s and ending in the transition $\delta(q_{n-1}, 0) = q_0$.
\end{itemize}
After some number of cycles through $q_0$, we eventually
arrive at $q_{n-1}$.  Letting $i$ denote the number of times a path of
type (b) is chosen (including the last path that arrives at $q_{n-1}$)
and $j$ denote the number of times a path of type (a) is chosen, we see
that the number of $1$'s in any accepted word must be of the form
$i(m-1) + jm$, with $i > 0$, $j \geq 0$.  The number of $0$'s along
such a path is then at least $i(n-m+1) - 1$, with the $-1$ in this expression
arising from the fact that the last part of the path terminates at $q_{n-1}$
without taking an additional $0$ transition back to $q_0$.

Thus
\begin{align*}
L(M_2) \subseteq \{ x \in \Sigma^* : \exists i,j \in {\mathbb N}, \text{ such that } i > 0, j \geq 0, \text{ and } \\
|x|_1  = i(m-1) + jm, \ |x|_0 \geq i(n-m+1)-1 \}.
\end{align*}
Furthermore,
for every $i,j \in {\mathbb N}, \text{ such that } i > 0, j \geq 0$,
there exists an $x \in L(M_2)$ such that $|x|_1  = i(m-1) + jm$, and
$|x|_0 = i(n-m+1)-1$.   This is obtained, for example, by cycling 
$j$ times from $q_0$ to $q_{m-1}$
and then back to $q_0$ via a transition on $1$, then $j-1$ times
from $q_0$ to $q_{n-1}$ and then back to $q_0$ via a transition on $0$,
and finally one more time from $q_0$ to $q_{n-1}$.  

It follows then that 
\begin{align*}
L(M_1 \cap M_2) &\subseteq \{ x \in \Sigma^*: \exists i,j \in {\mathbb N}, \text{ such that } i > 0, j \geq 0, \text{ and } \\
& |x|_1  = i(m-1) + jm, \  |x|_0 \geq i(n-m+1)-1\\
& \text{and } i(m-1) + jm \equiv 0\!\! \pmod {m} \}.%j(m) \equiv 0 \pmod {m} \}. %\equiv 0 ({\hbox{mod}} m) \}. 
\end{align*}
Further, for every such $i$ and $j$, there exists a corresponding element in $L(M_1 \cap M_2)$. Since $m-1$ and $m$ are relatively prime, the shortest such word corresponds to $i = m$, $j = 0$,  and satisfies $|x|_0 = m(n-m+1)-1$. In particular, a shortest accepted word is $(1^{m-1}0^{n-m+1})^{m-1}1^{m-1}0^{n-m}$, which is of length $mn-1$.
\ \ \ \qed

\bigskip

It is natural to try to extend the construction to an arbitrary number of
DFAs. However, we have found empirically that, over a two-letter alphabet,
the corresponding bound $mnp-1$ for three DFA's does not always hold.  For
example, there are no DFA's of $2, 2, $ and $3$ states for which the
shortest word in the intersection is of length $2\cdot 2\cdot 3 - 1$.

\end{document}

%% file: intersectm3.pstex_t
\begin{picture}(0,0)%
\epsfig{file=intersectm3.pstex}%
\end{picture}%
\setlength{\unitlength}{3947sp}%
\begingroup\makeatletter\ifx\SetFigFont\undefined%
\gdef\SetFigFont#1#2#3#4#5{%
  \reset@font\fontsize{#1}{#2pt}%
  \fontfamily{#3}\fontseries{#4}\fontshape{#5}%
  \selectfont}%
\fi\endgroup%
\begin{picture}(8455,3628)(1639,-3104)
\put(2401,389){\makebox(0,0)[lb]{\smash{\SetFigFont{12}{14.4}{\rmdefault}{\mddefault}{\updefault}0}}}
\put(4276,389){\makebox(0,0)[lb]{\smash{\SetFigFont{12}{14.4}{\rmdefault}{\mddefault}{\updefault}0}}}
\put(9826,389){\makebox(0,0)[lb]{\smash{\SetFigFont{12}{14.4}{\rmdefault}{\mddefault}{\updefault}0}}}
\put(3151,-211){\makebox(0,0)[lb]{\smash{\SetFigFont{12}{14.4}{\rmdefault}{\mddefault}{\updefault}1}}}
\put(4951,-211){\makebox(0,0)[lb]{\smash{\SetFigFont{12}{14.4}{\rmdefault}{\mddefault}{\updefault}1}}}
\put(8626,-211){\makebox(0,0)[lb]{\smash{\SetFigFont{12}{14.4}{\rmdefault}{\mddefault}{\updefault}1}}}
\put(9826,-1486){\makebox(0,0)[lb]{\smash{\SetFigFont{12}{14.4}{\rmdefault}{\mddefault}{\updefault}1}}}
\put(7351,-1786){\makebox(0,0)[lb]{\smash{\SetFigFont{12}{14.4}{\rmdefault}{\mddefault}{\updefault}1}}}
\put(6976,-2236){\makebox(0,0)[lb]{\smash{\SetFigFont{12}{14.4}{\rmdefault}{\mddefault}{\updefault}1}}}
\put(8776,-2986){\makebox(0,0)[lb]{\smash{\SetFigFont{12}{14.4}{\rmdefault}{\mddefault}{\updefault}0}}}
\put(6901,-2911){\makebox(0,0)[lb]{\smash{\SetFigFont{12}{14.4}{\rmdefault}{\mddefault}{\updefault}0}}}
\put(3151,-2986){\makebox(0,0)[lb]{\smash{\SetFigFont{12}{14.4}{\rmdefault}{\mddefault}{\updefault}0}}}
\put(2026,-1561){\makebox(0,0)[lb]{\smash{\SetFigFont{12}{14.4}{\rmdefault}{\mddefault}{\updefault}0,1}}}
\put(2251,-361){\makebox(0,0)[lb]{\smash{\SetFigFont{12}{14.4}{\rmdefault}{\mddefault}{\updefault}$q_0$}}}
\put(2161,-2821){\makebox(0,0)[lb]{\smash{\SetFigFont{12}{14.4}{\rmdefault}{\mddefault}{\updefault}$q_{n-1}$}}}
\put(7666,-2836){\makebox(0,0)[lb]{\smash{\SetFigFont{12}{14.4}{\rmdefault}{\mddefault}{\updefault}$q_m$}}}
\put(9601,-2836){\makebox(0,0)[lb]{\smash{\SetFigFont{12}{14.4}{\rmdefault}{\mddefault}{\updefault}$q_{m-1}$}}}
\put(9601,-361){\makebox(0,0)[lb]{\smash{\SetFigFont{12}{14.4}{\rmdefault}{\mddefault}{\updefault}$q_{m-2}$}}}
\put(4111,-346){\makebox(0,0)[lb]{\smash{\SetFigFont{12}{14.4}{\rmdefault}{\mddefault}{\updefault}$q_1$}}}
\end{picture}